\documentclass[aps,prb,twocolumn,floatfix,showpacs]{revtex4}
\usepackage{amsmath,amssymb,graphicx,bm}
\begin{document}
\title{Integrability of the diffusion pole in the diagrammatic
  description of noninteracting electrons in a random potential}

\author{V.  Jani\v{s}}

\affiliation{Institute of Physics, Academy of Sciences of the Czech
  Republic, Na Slovance 2, CZ-18221 Praha, Czech Republic }
\email{janis@fzu.cz}

\date{\today}


\begin{abstract} We discuss restrictions on the existence of the
  diffusion pole in the translationally invariant diagrammatic
  treatment of disordered electron systems. We use the Bethe-Salpeter
  equations for the two-particle vertex in the electron-hole and the
  electron-electron scattering channels and derive for systems with
  time reversal symmetry a nonlinear integral equation the
  two-particle irreducible vertices from both channels must obey. We
  use this equation to test the existence of the diffusion pole in the
  two-particle vertex.  We find that a singularity of the diffusion
  pole can exist only if it is integrable, that is only in the
  metallic phase in dimensions $d>2$.
\end{abstract} \pacs{72.10.Bg, 72.15.Eb, 72.15.Qm} \maketitle 
%

\textbf{\textit{Introduction}}.
Scattering of free charge carriers on impurities and lattice
imperfections can lead at low-temperatures to a metal-semiconductor
transition. There are two qualitatively different scenarios how a
metal can turn insulating due to excessive scatterings on impurities.
In the first case the metal-insulator transition materializes in
substitutional alloys when charge carriers are expelled from the Fermi
surface and an energy gap develops. This transition, called split
band, is qualitatively well understood and quantitatively well modeled
by a mean-field solution.\cite{Elliot74,Gonis92} The second type of a
metal-insulator transition is much more complicated and up to now not
completely understood. Electrons in a metal with random impurities can
lose their ability to diffuse on macroscopic scales. Such scenario was
first suggested by Anderson\cite{Anderson58} and is now called
Anderson localization transition.

One of principle obstacles of full understanding of the Anderson
localization transition is inability to describe vanishing of
diffusion of electrons analytically even in its simplest model version
and reconcile results from analytic and numerical approaches.
Analytic, mostly diagrammatic and field-theoretic approaches in the
thermodynamic limit indicate that the critical behavior at the
Anderson localization transition fits the one-parameter scaling scheme
with a single correlation length controlling the long-range
fluctuations.\cite{Wegner76,Abrahams79} On the other hand, an
increasing number of numerical studies of the Anderson localization
transition in finite volumes suggest that instead of homogeneous,
translationally invariant parameters one has to take into
consideration distributions of conductances or local particle
densities.\cite{Shapiro90,Markos93} The two different methodological
approaches, analytic and numerical, disagree not only on the number of
relevant controlling parameters needed to understand Anderson
localization but also on the critical behavior and the values of the
critical exponents.\cite{Kramer93,Lee85} Neither of these approaches
is, however, absolutely conclusive in delivering ultimate answers.

In case of disagreement of results from two rather well established
and otherwise reliable methods one has to revisit the assumptions
under which either results were derived and to which restrictions they
are subject.  One of the most important features used in the
description of the critical behavior of the Anderson localization
transition is a singular low-energy behavior of the density-density
correlation function of disordered systems. This singularity has form
of a resolvent of a diffusion equation and is called the diffusion
pole. The existence of the diffusion pole and a connection of the
diffusion constant with conductivity are consequences of conservation
laws in random systems.\cite{Janis03a} Conservation laws should be a
firm part of any reliable theory. We, however, showed recently that an
asymptotic solution of the Anderson model of noninteracting electrons
in high spatial dimensions does not fully obey conservation of
probability.\cite{Janis05a} We suggested a qualitative explanation for
such an unexpected behavior but more importantly, we amassed arguments
that unrestricted compliance with the conservation law is in random
systems in conflict with analyticity of the spectral
function.\cite{Janis04a,Janis04b} Since discussion about the form of
the diffusion pole is still ongoing,\cite{Suslov06,Brndiar08} we trace
down in this paper the origin and set exact restrictions on the form
of the diffusion pole derived within the translationally invariant
description of disordered systems in the thermodynamic limit. We first
thoroughly analyze the assumptions used to derive the diffusion pole
and then prove an assertion about the acceptable form of this
singularity without referring or resorting to any specific
approximation. We find that in systems invariant with respect to time
inversion the diffusion pole must be integrable in momentum space.

\textbf{\textit{Definitions and assumptions}}.
We model the system of non-interacting electrons by a lattice gas
described by an Anderson Hamiltonian\cite{Anderson58}
\begin{equation}\label{eq:Anderson-Hamiltonian}
  \widehat{H}=\sum_{\mathbf{k}}
  |\mathbf{k}\rangle \epsilon(\mathbf{k}) \langle\mathbf{k}|
  + \sum_i |i\rangle V_i \langle i|
\end{equation}
used to capture the impact of randomness on the electronic structure
of metallic alloys as well as to understand vanishing of diffusion in
the limit of strong randomness. The first, homogeneous, part of this
Hamiltonian is kinetic energy and is diagonalized in momentum space
(Bloch waves). The second sum runs over lattice sites and describes a
site-diagonal random potential. Values $V_i$ at different positions
are uncorrelated and follow a probability distribution $P(V_i)$. This
term is diagonalized in the direct space by local Wannier states. The
two operators do not commute, quantum fluctuations become important
and the full Anderson Hamiltonian cannot be easily diagonalized. The
only way to keep analytic control of the behavior of equilibrium
states of the Anderson model is to go directly to the thermodynamic
limit. Standardly it is approached by applying the ergodic theorem,
that is, summation over lattice sites equals the configurational
averaging. This means that we assume self-averaging property for all
quantities of interest. This need not be, however, always fulfilled as
we know from studies of Anderson localization. Presently we disregard
this option from consideration as well as the problem of the existence
of the thermodynamic limit.

Ergodicity itself, however, does not simplify the process of averaging
over randomness. Another assumption must be adopted to master this
problem. We assume that the thermodynamic limit can be performed
independently term by term in the expansion in powers of the random
potential. It means that we expect that the configurationally averaged
perturbation expansion in the random potential converges for all
quantities of interest.

Thermodynamic limit has an important simplifying consequence for
macroscopic (averaged) quantities.  The spectrum of a random
Hamiltonian in the thermodynamic limit is invariant with respect to
lattice translations. It means that operators $\widehat{H}$ and
$\widehat{T}_{\mathbf{R}}^{\phantom{\dagger}}\widehat{H}
\widehat{T}_{\mathbf{R}}^\dagger$, where $\widehat{T}_{\mathbf{R}}$ is
the operator of translation with a lattice vector $\mathbf{R}$, have
identical spectrum of eigenvalues with translationally shifted
eigenvectors. A lattice translation by a vector $\mathbf{R}_n$ applied
to the Anderson Hamiltonian from Eq.~\eqref{eq:Anderson-Hamiltonian}
generates a new one, $\sum_{\mathbf{k}} |\mathbf{k}\rangle
\epsilon(\mathbf{k}) \langle\mathbf{k}| + \sum_i |i + n\rangle V_{i}
\langle i + n|$ having the same distribution of random energies.
Unless we break translational symmetry in thermodynamic states, we are
unable to distinguish translationally shifted Hamiltonians. We cannot,
however, break translational invariance of the thermodynamic states
arbitrarily, since their symmetry should be in concord with the
spatial distribution of the eigenstates of the Hamiltonian for the
given configuration of the random potential. Since we do not know this
spectrum, we must treat all lattice translations of the Hamiltonian as
equivalent and instead of one Hamiltonian we are able to describe only
the whole class of equivalent Hamiltonians $\widehat{T}_{\mathbf{R}}
^{\phantom{\dagger}}\widehat{H} \widehat{T}_{\mathbf{R}}^\dagger$. In
this way we cannot distinguish directly between extended and localized
eigenstates of the random potential, since the localized states are
represented by a class of vectors differing by lattice translations.
 
The natural basis for translationally invariant quantities is formed
by Bloch waves labeled by quasimomenta. We generically denote
$\mathbf{k}, \mathbf{q}$ fermionic and bosonic (transferred) momenta
respectively.  The fundamental building blocks of the translationally
invariant description of disordered electrons are averaged one- and
two-particle resolvents $G(\mathbf{k},z)$ and $G^{(2)}_{{\bf k}{\bf
    k}'}(z_+,z_-;\mathbf{q})$, where $z_+ = E + \omega + i\eta$ and
$z_- = E - i\eta$ are complex energies with $E$ standing for the Fermi
energy, $\omega$ for the bosonic transfer frequency (energy), and
$\eta$ is a (infinitesimally) small damping (convergence) factor. We
adopt the electron-hole representation for the two-particle Green
function with $\mathbf{k}$ and $\mathbf{k}'$ for incoming and outgoing
electron momenta.  The bosonic momentum $\mathbf{q}$ measures the
difference between the incoming momenta of the electron and the
hole. Energies of the electron and the hole $z_+,z_-$ in systems with
noninteracting particles are external parameters.

The averaged one-electron resolvent in disordered systems can be
represented as in many-body theories via an irreducible vertex -- the
self-energy $\Sigma(\mathbf{k},z)$. We can write a Dyson equation for
it
\begin{equation}\label{eq:Dyson}
  \left\langle \left\langle\mathbf{k}\left| \frac 1{z\widehat{1} -
          \widehat{H}}\right| \mathbf{k}'\right\rangle \right\rangle_{av}  =\frac
  {\delta(\mathbf{k} - \mathbf{k}')}{z - \epsilon(\mathbf{k}) -
    \Sigma(\mathbf{k},z)}\,.
\end{equation}
The self-energy $\Sigma(\mathbf{k},z)$ stands for the impact of the
scatterings of the electron on random impurities. Knowledge of the
self-energy is then sufficient to determine the energy spectrum,
spectral density and in general all aspects of propagation of single
particles in disordered media.

The two-particle resolvent $G^{(2)}$ can then be represented via a
two-particle vertex $\Gamma$ defined from an equation
\begin{multline}\label{eq:G2-Gamma}
G^{(2)}_{{\bf k}{\bf
      k}'}(z_+,z_-;\mathbf{q}) =\\
  \left\langle \left\langle \mathbf{q} + \mathbf{k},\mathbf{k} \left|
        \frac 1{z_+ - \widehat{H}}\otimes \frac 1{z_- -
          \widehat{H}}\right| \mathbf{k}', \mathbf{q} + \mathbf{k}'
    \right\rangle \right\rangle_{av} \\ \equiv \left\langle
    \left\langle\mathbf{k} \left| \frac 1{z_+ -
          \widehat{H}}\right|\mathbf{k}' \right\rangle \left\langle
      \mathbf{q} + \mathbf{k}' \left| \frac 1{z_- -
          \widehat{H}}\right| \mathbf{q} + \mathbf{k}\right\rangle
  \right\rangle_{av} \\ =  G(\mathbf{k}, z_+)G( \mathbf{q} +
  \mathbf{k}, z_-)\left[\delta(\mathbf{k} - \mathbf{k}') \right. \\
  \left. + \ \Gamma_{{\bf k}{\bf k}'}(z_+,z_-;\mathbf{q})
    G(\mathbf{k}',z_+) G( \mathbf{q} +
    \mathbf{k}',z_-)\right]\ \end{multline} where $\otimes$ denotes
the direct product of operators.  The two-particle vertex introduces a
disorder-induced correlation into the two-particle propagation.
Analogously to the self-energy it measures the net impact of
scatterings on impurities on the motion of particles in the presence
of other particles.

The two-particle vertex $\Gamma$ can further be simplified by
introducing an irreducible vertex $\Lambda$ playing the role of a
two-particle self-energy. The irreducible and the full vertex are
connected by a Bethe-Salpeter equation.  Unlike the one-particle
irreducibility, the two-particle irreducibility is
ambiguous.\cite{Janis01b} There are two types of two-particle
irreducibility in systems with elastic scatterings only, electron-hole
and electron-electron. They are characterized by different
Bethe-Salpeter equations.  The Bethe-Salpeter equation in the
electron-hole scattering channel then reads
\begin{subequations}\label{eq:BS}
  \begin{multline}\label{eq:BS-eh}
    \Gamma_{\mathbf{k}\mathbf{k}'}(\mathbf{q}) =
    \Lambda^{eh}_{\mathbf{k}\mathbf{k}'}(\mathbf{q})\\ + \frac
    1N\sum_{\mathbf{k}''}
    \Lambda^{eh}_{\mathbf{k}\mathbf{k}''}(\mathbf{q})
    G_+(\mathbf{k}'') G_-( \mathbf{q} + \mathbf{k}'')
    \Gamma_{\mathbf{k}''\mathbf{k}'}(\mathbf{q})\ . \end{multline}
  We suppressed the frequency variables in Eq.~\eqref{eq:BS-eh}, since
  they are not dynamical ones. They can be easily deduced from the
  one-electron propagators $G_\pm(\mathbf{k})= G^{R,A}(\mathbf{k})
  \equiv G(\mathbf{k},z_\pm)$ used there.

  We can introduce another nonequivalent representation of the
  two-particle vertex. If we sum explicitly multiple scatterings of
  two electrons (holes) we can construct an alternative Bethe-Salpeter
  equation\cite{Janis01b}
  \begin{multline}\label{eq:BS-ee}
    \Gamma_{\mathbf{k}\mathbf{k}'}(\mathbf{q}) =
    \Lambda^{ee}_{\mathbf{k}\mathbf{k}'}(\mathbf{q}) + \frac
    1N\sum_{\mathbf{k}''}
    \Lambda^{ee}_{\mathbf{k}\mathbf{k}''}(\mathbf{q} + \mathbf{k}' -
    \mathbf{k}'')\\ \times G_+(\mathbf{k}") G_-(\mathbf{q} +
    \mathbf{k} + \mathbf{k}' - \mathbf{k}'')
    \Gamma_{\mathbf{k}''\mathbf{k}'}(\mathbf{q} + \mathbf{k} -
    \mathbf{k}'')\ . \end{multline}\end{subequations}%
We introduced an irreducible vertex in the electron-electron
scattering channel $\Lambda^{ee}$.  Irreducible vertices
$\Lambda^{eh}$ and $\Lambda^{ee}$ do not include isolated pair
electron-hole and electron-electron scatterings, respectively.

\textbf{\textit{Diffusion pole and electron-hole symmetry}}.
Noninteracting particles scattered on impurities are marked by a
diffusion pole.  The low-energy limit of a special matrix element of
the two-particle resolvent, electron-hole correlation function, has
the following asymptotics for $q \to 0$ and $ \omega/q \to 0$
\begin{align}
  \label{eq:Phi-diffusion}
  \Phi^{RA}_{E}({\bf q},\omega) &= \frac 1{N^2}\sum_{\mathbf{k}\mathbf
    {k'}}G^{RA}_{{\bf k}{\bf k}'}(E + \omega, E; \mathbf{q})    \nonumber\\
  &\doteq \frac {2\pi n_F}{-i\omega + D(\omega)q^2} + O(q^0,\omega^0)
\end{align}
where $n_F$ is the density of one-particle states at the Fermi
level.\cite{Janis03a} We used an abbreviation for the energy arguments
$G^{RA}_{{\bf k}{\bf k}'}(E + \omega, E; \mathbf{q}) \equiv
G^{(2)}_{{\bf k}{\bf k}'}(E + \omega + i0^+, E- i0^+;\mathbf{q})$. The
low-energy electron-hole correlation function becomes a propagator of
a diffusion equation.

Such a low-energy behavior is not evident and to prove it one has to
use Ward identities connecting one- and two-particle averaged
functions.  Ward identities reflect conservation laws.  In disordered
noninteracting systems we have probability (mass or charge)
conservation. It is mathematically equivalent to completeness of the
Hilbert space of Bloch waves. First Ward identity due to charge
conservation was derived for disordered systems within the mean-field
approximation by Velick\'y\cite{Velicky69} and later extended beyond
this approximation in Ref.~\onlinecite{Janis01b}. It is a consequence
of an operator identity
\begin{equation}\label{eq:Ward-operator}
  \frac 1{z_+ -\widehat{H}} \ \frac 1{z_- -\widehat{H}} = \frac 1{z_- - z_+}
  \left[ \frac 1{z_+ -\widehat{H}} - \frac 1{z_- -\widehat{H}}\right]
\end{equation}
where the multiplication is the standard operator (matrix) one. This
identity holds for any one-particle Hamiltonian.  In the thermodynamic
limit we must, however, average this identity and the averaging
procedure need not conserve all its aspects when projected onto
translationally invariant states.\cite{Janis04a} When using the above
identity in the evaluation of the homogeneous part of the
electron-hole correlation function, that is $q=0$, we obtain
\begin{equation}
  \label{eq:W_BV_momentum}
  \Phi^{RA}_E(\mathbf{0},\omega) \doteq  \frac {2\pi
    n_F}{-i\omega}\, .
\end{equation}
No spatial fluctuations ($q\neq 0$) of the correlation function in the
low-frequency limit can be deduced from the Velick\'y-Ward
identity. To derive the spatial behavior of the diffusion pole in
Eq.~\eqref{eq:Phi-diffusion} one has to resort to another relation
introduced by Vollhardt and W\"olfle.\cite{Vollhardt80a} It utilizes
the Dyson and Bethe-Salpeter equations, Eq.~\eqref{eq:Dyson} and
Eq.~\eqref{eq:BS-eh}, and relates the one- and two-particle
irreducible functions $\Sigma$ and $\Lambda^{eh}$, respectively. It
reads
\begin{multline} \label{eq:VWW-identity} \Sigma^R( \mathbf{q} + {\bf
    k} ,E + \omega) - \Sigma^A({\bf k},E) = \frac 1N \sum_{{\bf
      k}'}\Lambda^{RA}_{{\bf k}{\bf k}'}(E + \omega, E; \mathbf{q})\\
  \times \left[ G^R( \mathbf{q} + {\bf k}' ,E + \omega) - G^A({\bf
      k}',E) \right] \ . \end{multline}
and was proved diagrammatically (perturbatively).  Using the
Bethe-Salpeter equation one can show that in the homogeneous limit
$q=0$ this identity reflects the continuity equation and hence is
equivalent to the Velick\'y-Ward identity.
Equation~\eqref{eq:VWW-identity} together with the Bethe-Salpeter
equation are then used to show that the long-distance fluctuations of
the low-energy limit of the correlation function are controlled by a
diffusion constant.  Introducing a dynamical diffusion constant
$D(\omega)$ we can represent the full leading low-energy asymptotics
of the electron-hole correlation function as in
Eq.~\eqref{eq:Phi-diffusion}.\cite{Janis03a} Note that
identity~\eqref{eq:W_BV_momentum} holds for both pure and random
systems, the actual diffusion pole, however, is only the singularity
from Eq.~\eqref{eq:Phi-diffusion} with the momentum dependence of the
low-energy behavior.  To prove such a spatially diffusive behavior the
Bethe-Salpeter equation becomes an indispensable tool.

Another important feature of noninteracting electrons on a bipartite
lattice without external magnetic field and spin-orbit coupling is the
time reversal symmetry. Time inversion is equivalent to reversing the
direction of the particle propagation, that is
$\mathbf{k}\to-\mathbf{k}$. The electron and the hole interchange
their roles. The time-reversal invariance for the one-particle
propagator then means $ G(\mathbf{k},z) = G(-\mathbf{k},z)$. Time
inversion leads to nontrivial symmetries when applied onto one of the
fermion propagators in two-particle functions. The electron-hole
transformation can be represented either by reversing the electron
line leading to a transformation $\mathbf{k} \to -\mathbf{k}',
\mathbf{k}' \to -\mathbf{k}, \mathbf{q} \to \mathbf{Q}$ or by
reversing the hole propagator $\mathbf{k} \to \mathbf{k}, \mathbf{k}'
\to \mathbf{k}', \mathbf{q} \to -\mathbf{Q}$ for the electron-hole
function. Here we denoted $ \mathbf{Q}= \mathbf{q} + \mathbf{k} +
\mathbf{k}'$.  We then obtain two symmetry relations for the full
two-particle vertex
\begin{subequations}\label{eq:EH-Symmetry}
  \begin{align}\label{eq:EH-Symmetry-full}
    \Gamma_{\mathbf{k}\mathbf{k}'}(\mathbf{q})& =
    \Gamma_{-\mathbf{k}'-\mathbf{k}}( \mathbf{Q}) =
    \Gamma_{\mathbf{k}\mathbf{k}'}( -\mathbf{Q}).
  \end{align}
  The two-particle irreducible vertices are not invariant with respect
  to time inversion, since the electron-hole vertex is transformed
  onto the electron-electron one and vice versa. We then have the
  following electron-hole symmetry relations
  \begin{align}\label{eq:EH-Symmetry-irreducible}
    \Lambda^{ee}_{\mathbf{k}\mathbf{k}'}(\mathbf{q}) &
    =\Lambda^{eh}_{-\mathbf{k}'-\mathbf{k}}(\mathbf{Q}) =
    \Lambda^{eh}_{\mathbf{k}\mathbf{k}'}(-\mathbf{Q}).
  \end{align}\end{subequations}
This relation says that Bethe-Salpeter equation~\eqref{eq:BS-eh}
transforms upon time inversion in one particle line onto
Bethe-Salpeter equation~\eqref{eq:BS-ee}.  When the invariance with
respect to the electron-hole transformation is applied to the
correlation function we obtain
\begin{equation}\label{eq:Phi-aternative}
  \Phi^{RA}_E(\mathbf{q}, \omega) =
  \frac 1{N^2}\sum_{\mathbf{k}\mathbf{k}'} G^{RA}_{\mathbf{k}\mathbf{k}'}(E+
  \omega, E ;-\mathbf{q}- \mathbf{k}- \mathbf{k}')\ .
\end{equation}
This representation together with the Ward identity,
Eq.~\eqref{eq:VWW-identity}, tell us that the same low-energy
singularity for $\omega,\mathbf{q}^2 \to 0$ must emerge with the same
weight in the averaged two-particle resolvent
$G^{RA}_{\mathbf{k}\mathbf{k}'}(E+\omega,E; \mathbf{q})$ also in the
limit $\omega,(\mathbf{k} + \mathbf{k}' + \mathbf{q})^2 \to 0$.

The uncorrelated propagation of electrons in a random potential does
not contain the diffusion pole, and hence it must emerge in the vertex
function $\Gamma$.  Taking into account the time-reversal invariance
we can single out the singular parts of the electron-hole symmetric
two-particle vertex and obtain
\begin{multline}\label{eq:Gamma-poles}
  \Gamma^{RA}_{\mathbf{k}\mathbf{k}'}(\mathbf{q}, \omega) =
  \gamma^{RA}_{\mathbf{k}\mathbf{k}'}(\mathbf{q},\omega) +
  \frac{\varphi^{RA}_{\mathbf{k}\mathbf{k}'}}{-i\omega +
    D(\omega)\mathbf{q}^2}\\ +
  \frac{\varphi^{RA}_{\mathbf{k}\mathbf{k}'}} {-i\omega +
    D(\omega)(\mathbf{q} + \mathbf{k} + \mathbf{k}')^2} \ .
\end{multline}
The reduced vertex $\gamma^{RA}$ has a marginal and thermodynamically
irrelevant singularity for $\omega\to0$ at $k= k' = q =0$. It can,
nevertheless, display another singular behavior in fermionic variables
$\mathbf{k},\mathbf{k}'$ that is not derivable from the diffusion pole.
Such a singularity must not, however, affect the form of the diffusion
pole in the electron-hole correlation function for $q \to 0$. The second
term on the right-hand side of Eq.~\eqref{eq:Gamma-poles} dominates in the
leading order of the limit $q \to 0$, $\omega \to 0$ while the third one
in the limit $\mathbf{q} + \mathbf{k} + \mathbf{k}' \to 0$, $\omega \to
0$. We used the dynamical form of the diffusion constant $D(\omega)$ so
that the localization phase would fit. Equation~\eqref{eq:Gamma-poles} is
the most general form of the two-particle vertex reproducing the diffusion
pole in the correlation function $\Phi$. The singularity for
$\mathbf{q}\to 0$ is the diffusion pole while the other for $\mathbf{q} +
\mathbf{k} + \mathbf{k}' \to 0$ is the Cooper pole caused by multiple
electron-electron scatterings. To conform this representation with
Eq.~\eqref{eq:W_BV_momentum} we have to satisfy a normalization condition
that in the metallic phase ($D(0) >0$) reads
\begin{align}\label{eq:Phi-consistency}
  \frac 1{N^{2}}\sum_{\mathbf{k}\mathbf{k}'}
  |G_+(\mathbf{k})|^2\varphi^{RA}_{\mathbf{k}\mathbf{k}'}
  |G_+(\mathbf{k}')|^2 & = 2\pi n_F\, .
\end{align}

\textbf{\textit{Parquet equations with time-reversal symmetry}}. %
The full two-particle vertex symmetric with respect to the electron-hole
transformation  can alternatively be decomposed by
means of the so-called parquet equation that can be represented in
various equivalent ways
\cite{Janis01b}
\begin{align}\label{eq:parquet-equation}
  \Gamma_{\mathbf{k}\mathbf{k}'}(\mathbf{q}) &=
  \Lambda^{ee}_{\mathbf{k}\mathbf{k}'}(\mathbf{q}) +
  \mathcal{K}^{ee}_{\mathbf{k}\mathbf{k}'}(\mathbf{q}) =
  \Lambda^{eh}_{\mathbf{k}\mathbf{k}'}(\mathbf{q}) +
  \mathcal{K}^{eh}_{\mathbf{k}\mathbf{k}'}(\mathbf{q})    \nonumber\\
  & = \mathcal{K}^{eh}_{\mathbf{k}\mathbf{k}'}(\mathbf{q}) +
  \mathcal{K}^{ee}_{\mathbf{k}\mathbf{k}'}(\mathbf{q}) +
  \mathcal{I}_{\mathbf{k}\mathbf{k}'}(\mathbf{q}) \nonumber \\ & =
  \Lambda^{ee}_{\mathbf{k}\mathbf{k}'}(\mathbf{q}) +
  \Lambda^{eh}_{\mathbf{k}\mathbf{k}'}(\mathbf{q}) -
  \mathcal{I}_{\mathbf{k}\mathbf{k}'}(\mathbf{q})
\end{align}
where $\mathcal{K}^{eh}_{\mathbf{k}\mathbf{k}'}(\mathbf{q})$ and
$\mathcal{K}^{ee}_{\mathbf{k}\mathbf{k}'}(\mathbf{q})$ are
two-particle reducible vertices in the electron-hole and
electron-electron channels, respectively. We denoted $\mathcal{I} =
\Lambda^{eh} \cap \Lambda^{ee}$ a two-particle fully irreducible
vertex, that is, a vertex irreducible simultaneously for both the
electron-hole and the electron-electron pair propagation (multiple
scatterings).

The parquet equations hold for the systems where the electron-hole and
the electron-electron multiple scatterings are nonequivalent, that is,
the corresponding two-particle irreducibilities are unambiguous and
excluding definitions of diagrammatic contributions.  The concept of
the parquet theory based on nonequivalence of two-particle
irreducibility can at best be understood in terms of sets of diagrams
where addition of functions is represented by union of sets of
diagrams the functions stand for.  Nonequivalence of the electron-hole
and the electron-electron multiple scatterings means
$\mathcal{K}^{ee}\cap\mathcal{K}^{eh}=\emptyset$. We trivially have in
each $\alpha$-channel $\Lambda^\alpha\cap \mathcal{K}^\alpha =
\emptyset$.  Further on, we have $\Lambda^{eh} =
\Lambda^{eh}\cap\Gamma =
(\Lambda^{eh}\cap\Lambda^{ee})\cup(\Lambda^{eh}\cap\mathcal{K}^{ee})
\subset \mathcal{I}\cup\mathcal{K}^{ee}$. On the other hand,
$\mathcal{K}^{ee} = \mathcal{K}^{ee}\cap\Gamma =
(\mathcal{K}^{ee}\cap\Lambda^{eh})\cup(\mathcal{K}^{ee}\cap\mathcal{K}^{eh})
= \mathcal{K}^{ee}\cap\Lambda^{eh}$. Hence $\mathcal{K}^{ee}\subset
\Lambda^{eh}$. Combining the above two relations we obtain
$\Lambda^{eh} = \mathcal{I}\cup\mathcal{K}^{ee}$ from which we reach
the parquet representations via irreducible or reducible vertices in
Eq.~\eqref{eq:parquet-equation},
$\Lambda^{eh}\cup\Lambda^{ee}\setminus\mathcal{I} =
\mathcal{I}\cup\mathcal{K}^{ee}\cup\mathcal{K}^{eh} = \Gamma$.

One must be careful when using the parquet decomposition for
noninteracting electrons with elastic scatterings only.  In this case
multiple scatterings on a single site are identical for both
channels. Hence, the two Bethe-Salpeter equations~\eqref{eq:BS} are
identical, when the one-electron propagators are purely local. We then
obtain $\Lambda^{eh} = \Lambda^{ee} = \mathcal{I}$. It means that
irreducible and reducible local diagrams coincide and the concept of
two-particle irreducibility becomes ambiguous.  To amend this problem
we introduce a stronger \textit{full two-particle irreducibility}
including also local scatterings where the electron and the hole are
indistinguishable. We denote this vertex $\mathcal{J}$. The
irreducible vertices $\mathcal{I},\Lambda^{eh}$ and $\Lambda^{ee}$ for
noninteracting electrons are then transformed in parquet
equations~\eqref{eq:parquet-equation} to
\begin{subequations}\label{eq:2PIR}
  \begin{align}\label{eq:2PIR-full}
    \mathcal{I}_{\mathbf{k}\mathbf{k}'}(\mathbf{q}) &=
    \mathcal{J}_{\mathbf{k}\mathbf{k}'}(\mathbf{q}) +
    \frac{\mathcal{J}^0 G_+ G_-}{1 - \mathcal{J}^0G_+ G_-}
    \mathcal{J}^0\
    , \\
    \Lambda^\alpha_{\mathbf{k}\mathbf{k}'}(\mathbf{q}) & =
    \overline{\Lambda}^\alpha_{\mathbf{k}\mathbf{k}'}(\mathbf{q}) +
    \frac{\mathcal{J}^0 G_+ G_-}{1 - \mathcal{J}^0G_+ G_-}
    \mathcal{J}^0
  \end{align}
\end{subequations}%
where $\mathcal{J}^0 =
N^{-3}\sum_{\mathbf{k}\mathbf{k}'\mathbf{q}}\mathcal{J}_{\mathbf{k}
  \mathbf{k}'}(\mathbf{q})$ and $G_\pm = N^{-1}\sum_{\mathbf{k}}
G_\pm(\mathbf{k})$ are the appropriate local (momentum-independent)
parts.  Vertex
$\overline{\Lambda}^\alpha_{\mathbf{k}\mathbf{k}'}(\mathbf{q})$ is
irreducible in channel $\alpha$ but does not contain multiple
scatterings on the same site.  It is important that the fully
irreducible vertex $\mathcal{J}_{\mathbf{k}\mathbf{k}'}(\mathbf{q})$
contains only cumulant averaged powers of the random potential on the
same lattice site so that double counting is avoided.

We now use the symmetries from Eq.~\eqref{eq:EH-Symmetry} to replace
the two irreducible vertices by a single function. We define
\begin{align}\label{eq:2PIR-reduction}
  \Lambda_{\mathbf{k}\mathbf{k}'}(\mathbf{q}) &\equiv
  \Lambda^{ee}_{\mathbf{k}\mathbf{k}'}(\mathbf{q}) =
  \Lambda^{eh}_{\mathbf{k}\mathbf{k}'}(-\mathbf{q} - \mathbf{k} -
  \mathbf{k}' ) \ . \end{align}%
We use this definition in parquet equation~\eqref{eq:parquet-equation}
where we represent the full vertex by Bethe-Salpeter
equation~\eqref{eq:BS-eh}. We then obtain a fundamental equation for
the irreducible vertex
\begin{multline}\label{eq:Lambda-nonlinear}
  \Lambda_{\mathbf{k}\mathbf{k}'}(\mathbf{q}) =
  \mathcal{I}_{\mathbf{k}\mathbf{k}'}(\mathbf{q}) \\ + \frac 1N
  \sum_{\mathbf{k}"} \Lambda_{\mathbf{k}\mathbf{k}"}(-\mathbf{q} -
  \mathbf{k} - \mathbf{k}" ) G_+(\mathbf{k}") G_-(\mathbf{q} +
  \mathbf{k}") \\ \times
  \left[\Lambda_{\mathbf{k}"\mathbf{k}'}(\mathbf{q})
    +\Lambda_{\mathbf{k}"\mathbf{k}'}(- \mathbf{q}- \mathbf{k}" -
    \mathbf{k}' ) - \mathcal{I}_{\mathbf{k}"\mathbf{k}'}(\mathbf{q})
  \right] \ . \end{multline}
This is a nonlinear integral equation for vertex $\Lambda$ from an
input $\mathcal{I}$ that may have multiple solutions. We choose the
physical one by matching it to a perturbative solution reached by an
iterative procedure with an auxiliary coupling constant $\lambda$ and
a starting condition $\Lambda^{(0)} = \lambda \mathcal{I}$. The
iteration procedure for a fixed coupling constant $\lambda$ is
determined by a recursion formula
\begin{multline}\label{eq:Lambda-Iterative}
  \frac 1N \sum_{\mathbf{k}"} \bigg[\delta_{\mathbf{k}", \mathbf{k}' }
  - \Lambda^{(n-1)}_{\mathbf{k}\mathbf{k}"}(-\mathbf{q} - \mathbf{k} -
  \mathbf{k}") G_+(\mathbf{k}") \\ \times G_-(\mathbf{q} +
  \mathbf{k}")\bigg]
  \left(\Lambda^{(n)}_{\mathbf{k}"\mathbf{k}'}(\mathbf{q}) -
    \lambda \mathcal{I}_{\mathbf{k}"\mathbf{k}'}(\mathbf{q}) \right) \\
  = \frac 1N \sum_{\mathbf{k}"} \Lambda^{(n-1)}_{\mathbf{k}
    \mathbf{k}"}(-\mathbf{q}  -  \mathbf{k} - \mathbf{k}") G_+(\mathbf{k}") \\
  \times G_-(\mathbf{q} + \mathbf{k}") \Lambda^{(n-1)}_{\mathbf{k}"
    \mathbf{k}'}(-\mathbf{q} - \mathbf{k}" - \mathbf{k}')\
  . \end{multline}
In this way vertex $\Lambda = \Lambda^{(\infty)}$ is completely
determined from the input, the fully irreducible vertex $\lambda
\mathcal{I}$. A physical solution for $\lambda = 1$ is reached only if
the iteration procedure converges for $0< \lambda \ll 1 $ and the
result can analytically be continued to $\lambda = 1$. This
construction of the physical solution corresponds to the
linked-cluster expansion from many-particle physics.\cite{Mahan90} The
iteration scheme from Eq.~\eqref{eq:Lambda-Iterative} is the only
available way to reach a physical solution and hence its convergence
and analyticity are of principal importance for the diagrammatic
description of disordered systems.  Using Eqs.~\eqref{eq:2PIR} we can
rewrite the above equation to another one for the irreducible vertex
$\overline{\Lambda}$ determined from $\mathcal{J}$. The latter vertex
is the genuine independent input. Notice that in single-site theories
with local one-electron propagators we obtain a solution
$\overline{\Lambda} = \mathcal{J}^0$ to  Eq.\eqref{eq:Lambda-Iterative}.

Equation~\eqref{eq:Lambda-nonlinear} (alternatively
Eq.~\eqref{eq:Lambda-Iterative}) is a fundamental equation of motion
for the two-particle irreducible vertex being electron-hole symmetric.
The corresponding full two-particle vertex obeys simultaneously both
the Bethe-Salpeter equations in the electron-hole and the
electron-electron channels, Eqs.~\eqref{eq:BS} and the two equations
are not identical, that is
$\Lambda^{eh}_{\mathbf{k}\mathbf{k}'}(\mathbf{q}) \neq
\Lambda^{ee}_{\mathbf{k}\mathbf{k}'}(\mathbf{q})$.  Nonlinearity of
the fundamental equation for the irreducible vertex poses restrictions
on the admissible form of the singular behavior in its solutions.
Singularities in the full vertex $\Gamma$ emerge only via
singularities in the irreducible vertex $\Lambda$.

\textbf{\textit{Assertion}}. %
\textit{Two-particle vertex $\Gamma$ of noninteracting electrons in a
  random potential can be decomposed into irreducible vertices
  as \begin{align}\label{eq:Gamma-Lambda-I}
    \Gamma_{\mathbf{k}\mathbf{k}'}(\mathbf{q}) &=
    \Lambda_{\mathbf{k}\mathbf{k}'}(\mathbf{q}) +
    \Lambda_{\mathbf{k}\mathbf{k}'}(-\mathbf{q} - \mathbf{k} -
    \mathbf{k}' ) - \mathcal{I}_{\mathbf{k}\mathbf{k}'}(\mathbf{q})\
    , \end{align} %
  if electrons and holes are distinguishable (non-equivalent)
  quasiparticles and the system is invariant with respect to time
  inversion (electron-hole symmetric).  We denoted $\mathcal{I}$ the
  two-particle fully irreducible vertex. Irreducible vertex $\Lambda$
  obeys Eq.~\eqref{eq:Lambda-nonlinear}. The diffusion pole in the
  full two-particle vertex $\Gamma$ may materialize only if it appears
  in the irreducible vertex $\Lambda$. Consequently, the diffusion and
  Cooper poles from Eq.~\eqref{eq:Gamma-poles} can exist in $\Gamma$ only
  in the metallic phase in spatial dimensions $d>2$.}

\textbf{\textit{Proof}}. %
Equation~\eqref{eq:Gamma-Lambda-I} is  a direct consequence of parquet
equation~\eqref{eq:parquet-equation} where the electron-hole symmetry,
Eq.~\eqref{eq:EH-Symmetry}, is used. The parquet equation holds if
the electron and the hole are distinguishable quasiparticles via their multiple
mutual scatterings. That is, electron-electron and electron-hole
scatterings do not lead to identical results.

We need not find the most general form of low-energy ($\omega\to 0$)
singularities compliant with Eq.~\eqref{eq:Lambda-nonlinear} but
rather check whether and when singularities from
representation~\eqref{eq:Gamma-poles} can emerge in solutions of
Eq.~\eqref{eq:Lambda-nonlinear}.

Vertex $\Lambda_{\mathbf{k}\mathbf{k}'}(\mathbf{q})$ contains the
diffusion pole of the full vertex $\Gamma$,
$\Lambda^{eh}_{\mathbf{k}\mathbf{k}'}(\mathbf{q})$ the Cooper pole and
the fully irreducible vertex
$\mathcal{I}_{\mathbf{k}\mathbf{k}'}(\mathbf{q})$ is free of these
poles.  This conclusion follows from an alternative form of
Eq.~\eqref{eq:Lambda-nonlinear} %
\begin{multline}\label{eq:Lambda-Gamma}
  \Lambda_{\mathbf{k}\mathbf{k}'}(\mathbf{q}) =
  \mathcal{I}_{\mathbf{k}\mathbf{k}'}(\mathbf{q}) \\ + \frac 1N
  \sum_{\mathbf{k}"} \Gamma_{\mathbf{k}\mathbf{k}"}(\mathbf{q})
  G_+(\mathbf{k}") G_-(\mathbf{q} + \mathbf{k}")
  \Gamma_{\mathbf{k}"\mathbf{k}'}(\mathbf{q}) \\ - \frac 1N
  \sum_{\mathbf{k}"} \left[\Lambda_{\mathbf{k}\mathbf{k}"}(\mathbf{q})
    - \mathcal{I}_{\mathbf{k}\mathbf{k}"}(\mathbf{q}) \right] \\
  \times G_+(\mathbf{k}") G_-(\mathbf{q} + \mathbf{k}")
  \Gamma_{\mathbf{k}"\mathbf{k}'}(\mathbf{q}) \end{multline}
where we used the fundamental parquet
equation~\eqref{eq:parquet-equation} to represent the integral kernel
$\Lambda$. The electron-hole symmetry leads in the limit $q\to
0$ and $\omega \to 0$ to an equation for the complex conjugate of the full two-particle vertex %
\begin{align}
  \Gamma^{RA}_{\mathbf{k}\mathbf{k}'}(\mathbf{q},\omega)^* & =
  \Gamma^{RA}_{\mathbf{k}' + \mathbf{q}, \mathbf{k} +
    \mathbf{q}}(-\mathbf{q}, -\omega)\ \end{align}
that we use to evaluate the convolution of the diffusion poles from
the full vertex $\Gamma$ in the first sum on the right-hand side of
Eq.~\eqref {eq:Lambda-Gamma}. We obtain for $\mathbf{k} = \mathbf{k}'$
in the leading order of $q\to 0$ and $\omega \to 0$
\begin{multline*}
  \frac 1N\sum_{\mathbf{k}"}\Gamma^{RA}_{\mathbf{k}\mathbf{k}"}
  (\mathbf{q},\omega) G_+(\mathbf{k}")  G_-(\mathbf{q} + \mathbf{k}")
  \Gamma^{RA}_{\mathbf{k}"\mathbf{k}}(\mathbf{q},\omega) |\\
  \xrightarrow[q\to 0,\omega\to 0]{} \frac 1N \sum_{\mathbf{k}"}
  \frac{|\varphi^{RA}_{\mathbf{k}"\mathbf{k}}
    G_+(\mathbf{k}")|^2}{\omega^2 + D(\omega)^2q^4}\ . \end{multline*}
This squared diffusion pole must be compensated by the second sum on the
right-hand side of Eq.~\eqref {eq:Lambda-Gamma}. It means that the
diffusion pole must be completely contained in function
$\Lambda_{\mathbf{k} \mathbf{k}'}(\mathbf{q}) -
\mathcal{I}_{\mathbf{k}\mathbf{k}'}(\mathbf{q})
=\mathcal{K}^{eh}_{\mathbf{k}\mathbf{k}'}(\mathbf{q})$. From the
electron-hole symmetry we then obtain that the Cooper pole must
completely be contained in function
$\mathcal{K}^{ee}_{\mathbf{k}\mathbf{k}'}(\mathbf{q})$ and consequently
the sum of the diffusion and the Cooper poles from the full vertex
$\Gamma_{\mathbf{k}\mathbf{k}'}(\mathbf{q})$ in Eq.~\eqref{eq:Gamma-poles}
is already part of function $\Gamma_{\mathbf{k} \mathbf{k}'}(\mathbf{q})
-\mathcal{I}_{\mathbf{k}\mathbf{k}'} (\mathbf{q})$. The fully irreducible
vertex $\mathcal{I}_{\mathbf{k}\mathbf{k}'} (\mathbf{q})$ is hence free of
the diffusion and Cooper poles.

We discuss first the behavior of the diffusion pole in the metallic phase
with $D(0) = D >0$. When inserting the singular part of the two-particle
vertex due to the diffusion pole we obtain the leading singularity on the
left-hand side of Eq.~\eqref{eq:Lambda-Iterative}
\begin{subequations}\begin{multline}\label{eq:LHS-singularity}
    S^L_{\mathbf{k}\mathbf{k}'} (\mathbf{q},\omega)=-\ \frac1{-i\omega
      + D q^2} \\ \times \frac 1N \sum_{\mathbf{k}"}
    \frac{\varphi^{RA}_{\mathbf{k}\mathbf{k}" }
      \varphi^{RA}_{\mathbf{k}"\mathbf{k}'} G_+(\mathbf{k}")
      G_-(\mathbf{q} + \mathbf{k}")} {-i \omega + D (\mathbf{q} +
      \mathbf{k} + \mathbf{k}")^2} \end{multline}
  and on its right-hand side
  \begin{multline}\label{eq:RHS-singularity}
    S^R_{\mathbf{k}\mathbf{k}'} (\mathbf{q},\omega)= \frac 1N
    \sum_{\mathbf{k}"} \frac{\varphi^{RA}_{\mathbf{k}\mathbf{k}" }
      \varphi^{RA}_{\mathbf{k}"\mathbf{k}'}} {-i \omega + D
      (\mathbf{q} + \mathbf{k} + \mathbf{k}")^2}\\ \times \frac
    {G_+(\mathbf{k}") G_-(\mathbf{q} + \mathbf{k}")} {-i \omega + D
      (\mathbf{q} + \mathbf{k}' + \mathbf{k}")^2}\
    .  \end{multline}\end{subequations}
Since the singular term from Eq.~\eqref{eq:LHS-singularity} contains
the complete form of the diffusion pole, the sum over momenta must not
bring any new singular contribution in small frequencies and is of
order $O(\omega^0)$. To assess the low-frequency behavior ($\omega\to
0$) of the sum over momenta we equal external fermionic momenta
$\mathbf{k}' = \mathbf{k}$ and use an asymptotic representation for
the contribution from the singular part of the integrands
\begin{subequations}\label{eq:Singularity}
  \begin{align}\label{eq:Singularity-left}
    S^L_{\mathbf{k}\mathbf{k}} (\mathbf{q},\omega) \doteq &
    \frac{\varphi^{RA}_{\mathbf{k},-\mathbf{q} -\mathbf{k}}
      \varphi^{RA}_{-\mathbf{q} -\mathbf{k},\mathbf{k}}
      G^R(\mathbf{q} +\mathbf{k}) G^A(\mathbf{k})} {-i\omega + Dq^2}\nonumber \\
    & \times \frac 1N \sum_{\mathbf{k}"}^\kappa \frac{1} {-i\omega + D
      (\mathbf{q} + \mathbf{k} + \mathbf{k}")^2}\
    ,\\ \label{eq:Singularity-right} S^R_{\mathbf{k}\mathbf{k}}
    (\mathbf{q},\omega) \doteq & \varphi^{RA}_{\mathbf{k},-\mathbf{q}
      -\mathbf{k}} \varphi^{RA}_{-\mathbf{q}
      -\mathbf{k},\mathbf{k}}G^R(\mathbf{q} +\mathbf{k})
    G^A(\mathbf{k}) \nonumber\\ & \times \frac 1N
    \sum_{\mathbf{k}"}^\kappa \frac{1} {\left[-i\omega + D (\mathbf{q}
        + \mathbf{k} + \mathbf{k}")^2\right]^2}\ ,
  \end{align}\end{subequations}
where $\kappa$ is an appropriate momentum cut-off.  The two
expressions cannot be more divergent in the low-frequency limit as
$(-i\omega)^{-1}$ for any value of the external momenta $\mathbf{q}$
and $\mathbf{k}$. Due to the normalization condition,
Eq.~\eqref{eq:Phi-consistency}, we find to each vector $\mathbf{k}$ a
set (of measure one) of momenta $\mathbf{q}$ so that
$\varphi^{RA}_{\mathbf{k},-\mathbf{q} - \mathbf{k} } \neq 0$.  If the
homogeneous case, $q = 0$, falls into this set then from
Eq.~\eqref{eq:Singularity-left} we obtain integrability of the
diffusion pole. If not, then for $\varphi^{RA}_{\mathbf{k},-\mathbf{q}
  - \mathbf{k} } \neq 0$ we obtain $ S^L_{\mathbf{k}\mathbf{k}}
(\mathbf{q},\omega) \propto (-i\omega)^{d/2 - 1}/(-i\omega + Dq^2)$
and $ S^R_{\mathbf{k}\mathbf{k}} (\mathbf{q},\omega) \propto
(-i\omega)^{d/2 - 2}$. For low dimensions $d\le 2$, both functions
$S^L_{\mathbf{k}\mathbf{k}} (\mathbf{q},\omega)$ and
$S^R_{\mathbf{k}\mathbf{k}} (\mathbf{q},\omega)$ have a stronger divergence
than $(-i\omega)^{-1}$  (for $q=0$)  and Eq.~\eqref{eq:Lambda-Iterative}
cannot be satisfied by any function $\Lambda_{\mathbf{k}\mathbf{k}}
(\mathbf{q},\omega)$. The diffusion pole can hence  exist in the metallic
phase only in dimensions $d>2$.

In the localized phase we expect the following low-energy asymptotics
($q \to 0, \omega \to 0$) of the dynamical diffusion
constant\cite{Vollhardt80b}
\begin{equation}\label{eq:diffusion-localized}
  \xi^2 = \frac{D(\omega)} {-i\omega} > 0
\end{equation}
where $\xi$ is a localization length.  Using this asymptotics we can
represent the singular part of the irreducible vertex $\Lambda$ as
follows
%
\begin{align}
  \label{eq:Lambda-localized} \Lambda^{sing}_{\mathbf{k}\mathbf{k}'}
  (\mathbf{q},\omega) &\doteq
  \frac{\varphi_{\mathbf{k}\mathbf{k}'}}{-i\omega} \ \frac{1}{1 +
    \xi^2\mathbf{q}^2}\ .
\end{align}
We utilize the electron-hole symmetry to evaluate the complex conjugate of
the irreducible vertex $\Lambda$ in the low-frequency $\omega \to 0$ and
momentum $q\to 0$ limit
\begin{align}
\Lambda^{RA}_{\mathbf{k}\mathbf{k}'}(\mathbf{q},\omega)^* & =
\Lambda^{RA}_{-\mathbf{k}-\mathbf{q}, -\mathbf{k}' -
\mathbf{q}}(\mathbf{q}, -\omega)\    \end{align}
and use it to derive  a condition
for vanishing of quadratic singularity of order
$\omega^{-2}$ on the right-hand side of
Eq.~\eqref{eq:Lambda-nonlinear}. After substituting
 the representation of vertex $\Lambda$  from
Eq.~\eqref{eq:Lambda-localized} and setting $\mathbf{k}'=\mathbf{k}$  and
$\mathbf{q}=0$ in Eq.~\eqref{eq:Lambda-nonlinear} we obtain
\begin{multline}\label{eq:Alpha-BS}
\frac 1N\sum_{\mathbf{k}''}
\left|\frac{\varphi^{RA}_{\mathbf{k}\mathbf{k}''} G_+(\mathbf{k}'')}{1 +
\xi^2(\mathbf{k} + \mathbf{k}'')^2}\right|^2\left[2 + \xi^2(\mathbf{k} +
\mathbf{k}'')^2 \right] =0\ .
\end{multline}%
This condition can be fulfilled only if the irreducible vertex
$\Lambda_{\mathbf{k}\mathbf{k}''}(\mathbf{q},\omega)$ is free of the
 singularity due to the diffusion pole for $q\to 0$ and $\omega\to 0$, that
is, $\varphi^{RA}_{\mathbf{k}\mathbf{k}'}=0$ point-wise. The diffusion pole
hence cannot exist in the localized phase.

\textbf{\textit{Discussion and conclusions}}.
The most severe consequence of the Assertion is nonexistence of the
diffusion pole in the localized phase in any dimension.  It means that
when approaching the low-energy limit $q \to 0$ and $\omega/q\to 0$ in
low dimensions ($d \le 2$) we cannot meet the diffusion-pole
singularity.  The localized phase must be reached in a non-critical or
a less critical manner than that of the diffusion pole. Theories, such
as the self-consistent theory of Anderson localization of Vollhardt
and W\"olfle,\cite{Vollhardt80b} leading to solutions with a
nonintegrable diffusion pole are in conflict with the Bethe-Salpeter
equation either in the electron-hole or in the electron-electron
channel or with the electron-hole symmetry at the two-particle level.

The Assertion poses no restriction on the expected form of the
diffusion pole in the metallic phase in dimensions $d >2$, since the
singularity is integrable.  The localized phase in $d >2$ is, however,
different. There the widely accepted behavior of the diffusion pole,
due to vanishing of the diffusion constant ($D=0$), becomes momentum
independent and hence nonintegrable. The fundamental
equation~\eqref{eq:Lambda-nonlinear} for the irreducible vertex
$\Lambda$ cannot lead to a two-particle vertex with such a
singularity. If the diffusion pole in $d > 2$ survives in the metallic
phase unchanged till the Anderson metal-insulator transition, there must
be a jump at the transition point at which the diffusion pole abruptly
ceases to exist.

Numerical simulations nevertheless seem to confirm the existence of the
diffusion pole in the localized phase.\cite{Brndiar08} There are two
possible conclusions we can draw from these incommensurable
results. One can speculate that some of the assumptions on which the
diagrammatic translationally invariant description of random systems
is based do not hold near to the Anderson localization
transition. Either ergodicity may be broken or one cannot average the
expansion in the random potential term by term, or the concept of
distinguishability of the electron-electron and electron-hole scatterings
is invalid. If this was true, then one had either to revisit the
derivation of the diffusion pole, being presently heavily based on the
Bethe-Salpeter representation of the two-particle vertex, or to question the
concept of electrons and holes as distinguishable quasiparticles in the
localized phase.

On the other hand,  numerical simulations are performable only on
rather small lattices where one cannot effectively reach the diffusive
regime $q\to 0$ with $\omega/q\to 0$.  Ref.~\onlinecite{Brndiar08}
investigates the opposite limit $\omega\to 0$ with $q/\omega\to 0$. As we
know,\cite{Janis03a} the two limits do not commute and the latter has
no relevance for the existence of the diffusion pole.  The numerically
observed $1/\omega$ behavior reflects only the Velick\'y
identity~\eqref{eq:W_BV_momentum} valid for  random as well as
pure systems. A conflict between the form of the diffusion pole and
the Bethe-Salpeter equation arises only in the critical region of the
latter. Ward identity~\eqref{eq:W_BV_momentum} cannot be extended
to inhomogeneous long-range fluctuations and the homogeneous
low-frequency limit $\omega\to 0$ with $q=0$ may be a singular point
having no macroscopic relevance for nonzero spatial fluctuations $q\ >
0$ in the thermodynamic limit.

Last but not least, we obtain as a consequence of
Eq.~\eqref{eq:Lambda-nonlinear} that the two-particle vertex
$\Gamma_{\mathbf{k}\mathbf{k}'}(\mathbf{q})$ in the metallic phase of
the most interesting spatial dimensions $2 < d < 4$ contains apart
from the diffusion and the Cooper pole also another low-energy
singularity for $\omega \to 0$ and $|\mathbf{k} - \mathbf{k}'| \to 0$.
We found that $S^R_{\mathbf{k}\mathbf{k}} (\mathbf{0},\omega)\doteq
(-i\omega)^{d/2 -2}$ in $d<4$ and hence a new singularity in vertex
$\Lambda^{RA}_{\mathbf{k}\mathbf{k}'} (\mathbf{q},\omega)$ emerges for
$\mathbf{k} - \mathbf{k}' \to 0$.  Due to the normalization condition,
Eq.~\eqref{eq:Phi-consistency}, it must be integrable, which is the
case for $d> 2$. This new singularity is compatible with the
decomposition from Eq.~\eqref{eq:Gamma-poles} of the two-particle
vertex $\Gamma$ into singularities caused by the diffusion pole. The
existence of a new singularity makes either the weight of the
diffusion pole $\varphi^{RA}_{\mathbf{k}\mathbf{k}'}$ or
$\gamma^{RA}_{\mathbf{k}\mathbf{k}'}(\mathbf{q},\omega)$ or both
singular with an integrable singularity. A new singularity in the
two-particle vertex indicates that the averaged two-particle functions
in spatial dimensions $2 < d < 4$ behave qualitatively differently and
have a richer analytic structure from those in higher dimensions. How
far this singularity influences the macroscopic behavior and transport
properties of disordered systems and in particular criticality of the
Anderson localization transition remains to be investigated.

To conclude, we proved in this paper that the diffusion pole in the
two-particle vertex can exist in the models of noninteracting electrons in
a random potential with time reversal symmetry only in the metallic phase
in dimensions $d > 2$. An equation of motion for the two-particle irreducible
vertices prevents the existence of the diffusion pole in the localized
phase. The existing translationally invariant descriptions of electrons in
a random potential predicting the existence of a pole in the localized
phase should hence be revisited. In view of our result, it seems very
difficult, if not impossible, to build up a consistent analytic theory of
Anderson localization with the diffusion pole in the localized phase.

\textbf{\textit{Acknowledgments}}. %
Research on this problem was carried out within project AVOZ10100520 of the
Academy of Sciences of the Czech Republic and supported in part by Grant
No. 202/07/0644 of the Grant Agency of the Czech Republic. I acknowledge
a fruitful collaboration and extensive discussions with J. Koloren\v c
on the problem of Anderson localization. I profited a lot particularly
from his critical remarks. I thank the Isaac Newton Institute for
Mathematical Sciences in Cambridge (UK) for hospitality extended to me
during my participation in the Programme \textit{Mathematics and Physics of
Anderson Localization}.


\begin{thebibliography}{99}
\bibitem{Elliot74}  R. J. Elliot, J. A. Krumhansl, and P. L. Leath, %
  \newblock Rev. Mod. Phys. {\bf 46}, 465 (1974).

\bibitem{Gonis92} A. Gonis, \newblock \textit{Green Functions for
    Ordered and Disordered Systems}(North Holland, Amsterdam, 1992).

\bibitem{Anderson58} P.~W.~Anderson, \newblock Phys. Rev. {\bf109},
  1492 (1958).

\bibitem{Wegner76} F. J. Wegner, \newblock Z. Physik B{\bf 35}, 327
  (1976).

\bibitem{Abrahams79} E. Abrahams, P. W. Anderson, D. C. Licciardello,
  and T. V. Ramakrishnan, \newblock Phys. Rev. Lett. {\bf 42}, 673
  (1979 .

\bibitem{Shapiro90} B. Shapiro, \newblock Phys. Rev. Lett. {\bf 65},
  1510 (1990).

\bibitem{Markos93} P. Marko\v s and B. Kramer, \newblock
  Phil. Mag. B{\bf 68}, 357 (1993).

\bibitem{Kramer93} B.~Kramer and A.~MacKinnon, \newblock Rep. Phys.
  {\bf 56}, 1469 (1993).
 
\bibitem{Lee85} P. A. Lee and R.V. Ramakrishnan, \newblock
  Rev. Mod. Phys.  {\bf 57}, 287 (1985).

\bibitem{Janis03a} V.~Jani\v s, J.~Koloren\v c, and V. \v Spi\v cka,
  \newblock Eur. Phys. J. B{\bf 35}, 77 (2003).
 
\bibitem{Janis05a} V.~Jani\v s and J.~Koloren\v c, \newblock
  Phys. Rev. B{\bf 71}, 033103, 245106 (2005).
 
\bibitem{Janis04a} V.~Jani\v s and J.~Koloren\v c, \newblock
  Phys. Stat. Sol. (b) {\bf 241}, 2032 (2004).
 
\bibitem{Janis04b} V.~Jani\v s and J.~Koloren\v c, \newblock
  Mod. Phys. Lett. B{\bf 18}, 1051 (2004).

\bibitem{Suslov06} I. M. Suslov, \newblock preprint
  arXiv:cond-mat/0612654.

\bibitem{Brndiar08} J. Brndiar and P. Marko\v s, \newblock
  Phys. Rev. B \textbf{77}, 115131 (2008).




\bibitem{Velicky69} B.~Velick\'y, \newblock Phys. Rev. {\bf 184}, 614
  (1969).
 
\bibitem{Janis01b} V.~Jani\v s, \newblock Phys. Rev. B{\bf 64}, 115115
  (2001).
 
\bibitem{Vollhardt80a} D. Vollhardt and P. W\"olfle, \newblock
  Phys. Rev.  Lett. {\bf 45}, 842 (1980).

\bibitem{Mahan90} G. D. Mahan, \newblock \textit{Many-Particle Physics},
Second Edition (Plenum Press, New York 1990).

\bibitem{Vollhardt80b} D.~Vollhardt and P. W\"olfle, \newblock
  Phys. Rev.  B{\bf 22}, 4666 (1980) and \newblock in
  \textit{Electronic Phase Transitions}, edited by W.~Hanke and
  Yu.~V.~Kopaev, (Elsevier Science Publishers B. V., Amsterdam, 1992).
\end{thebibliography}
\end{document}